\documentclass[aps,prl,twocolumn,superscriptaddress,showpacs,preprintnumbers,amsmath,amssymb]{revtex4}

\usepackage{graphicx}
\usepackage{dcolumn}
\usepackage{subfigure}
\usepackage{multirow}
\usepackage{comment}

\graphicspath{{ps}}

\begin{document}


\title{ \quad\\[1.0cm] Search for the decay $B^0\rightarrow DK^{*0}$ followed by $D\rightarrow K^-\pi^+$}



\affiliation{University of Bonn, Bonn}
\affiliation{Budker Institute of Nuclear Physics SB RAS and Novosibirsk State University, Novosibirsk 630090}
\affiliation{Faculty of Mathematics and Physics, Charles University, Prague}
\affiliation{University of Cincinnati, Cincinnati, Ohio 45221}
\affiliation{Department of Physics, Fu Jen Catholic University, Taipei}
\affiliation{Gifu University, Gifu}
\affiliation{Hanyang University, Seoul}
\affiliation{University of Hawaii, Honolulu, Hawaii 96822}
\affiliation{High Energy Accelerator Research Organization (KEK), Tsukuba}
\affiliation{Indian Institute of Technology Guwahati, Guwahati}
\affiliation{Indian Institute of Technology Madras, Madras}
\affiliation{Institute of High Energy Physics, Chinese Academy of Sciences, Beijing}
\affiliation{Institute of High Energy Physics, Vienna}
\affiliation{Institute of High Energy Physics, Protvino}
\affiliation{Institute for Theoretical and Experimental Physics, Moscow}
\affiliation{J. Stefan Institute, Ljubljana}
\affiliation{Kanagawa University, Yokohama}
\affiliation{Institut f\"ur Experimentelle Kernphysik, Karlsruher Institut f\"ur Technologie, Karlsruhe}
\affiliation{Korea Institute of Science and Technology Information, Daejeon}
\affiliation{Korea University, Seoul}
\affiliation{Kyungpook National University, Taegu}
\affiliation{\'Ecole Polytechnique F\'ed\'erale de Lausanne (EPFL), Lausanne}
\affiliation{Luther College, Decorah, Iowa 52101}
\affiliation{University of Maribor, Maribor}
\affiliation{Max-Planck-Institut f\"ur Physik, M\"unchen}
\affiliation{University of Melbourne, School of Physics, Victoria 3010}
\affiliation{Graduate School of Science, Nagoya University, Nagoya}
\affiliation{Kobayashi-Maskawa Institute, Nagoya University, Nagoya}
\affiliation{Nara Women's University, Nara}
\affiliation{National Central University, Chung-li}
\affiliation{National United University, Miao Li}
\affiliation{Department of Physics, National Taiwan University, Taipei}
\affiliation{Nippon Dental University, Niigata}
\affiliation{Niigata University, Niigata}
\affiliation{University of Nova Gorica, Nova Gorica}
\affiliation{Osaka City University, Osaka}
\affiliation{Pacific Northwest National Laboratory, Richland, Washington 99352}
\affiliation{Peking University, Beijing}
\affiliation{Research Center for Electron Photon Science, Tohoku University, Sendai}
\affiliation{Saga University, Saga}
\affiliation{University of Science and Technology of China, Hefei}
\affiliation{Seoul National University, Seoul}
\affiliation{Sungkyunkwan University, Suwon}
\affiliation{School of Physics, University of Sydney, NSW 2006}
\affiliation{Tata Institute of Fundamental Research, Mumbai}
\affiliation{Excellence Cluster Universe, Technische Universit\"at M\"unchen, Garching}
\affiliation{Toho University, Funabashi}
\affiliation{Tohoku Gakuin University, Tagajo}
\affiliation{Tohoku University, Sendai}
\affiliation{Department of Physics, University of Tokyo, Tokyo}
\affiliation{Tokyo Institute of Technology, Tokyo}
\affiliation{Tokyo Metropolitan University, Tokyo}
\affiliation{Tokyo University of Agriculture and Technology, Tokyo}
\affiliation{CNP, Virginia Polytechnic Institute and State University, Blacksburg, Virginia 24061}
\affiliation{Yamagata University, Yamagata}
\affiliation{Yonsei University, Seoul}

  \author{K.~Negishi}\affiliation{Tohoku University, Sendai} 
  \author{Y.~Horii}\affiliation{Kobayashi-Maskawa Institute, Nagoya University, Nagoya} 
  \author{Y.~Onuki}\affiliation{Department of Physics, University of Tokyo, Tokyo} 
  \author{T.~Sanuki}\affiliation{Tohoku University, Sendai} 
  \author{H.~Yamamoto}\affiliation{Tohoku University, Sendai} 

  \author{I.~Adachi}\affiliation{High Energy Accelerator Research Organization (KEK), Tsukuba} 
  \author{H.~Aihara}\affiliation{Department of Physics, University of Tokyo, Tokyo} 
 \author{D.~M.~Asner}\affiliation{Pacific Northwest National Laboratory, Richland, Washington 99352} 
  \author{T.~Aushev}\affiliation{Institute for Theoretical and Experimental Physics, Moscow} 
  \author{A.~M.~Bakich}\affiliation{School of Physics, University of Sydney, NSW 2006} 
% ...
\collaboration{The Belle Collaboration}

\begin{abstract}
We report a study of the decay $B^0\rightarrow D K^+\pi^-$ followed by $D\rightarrow K^-\pi^+$,
where $D$ indicates $D^0$ or $\bar{D}^0$.
We reconstruct the $D K^+\pi^-$ state in a phase space corresponding to $D K^{*}(892)^0$.
The $CP$-violating angle $\phi_3$ affects its decay rate via the interference between $b\rightarrow u$ and $b\rightarrow c$ transitions.
The result is obtained from a 711~${\rm fb}^{-1}$ data sample that contains 772~$\times 10^6~B\bar{B}$ pairs collected at the $\Upsilon(4S)$ resonance with the Belle detector at the KEKB asymmetric-energy $e^+ e^-$ collider.
We measure the ratio ${\cal R}_{DK^{*0}} \equiv \Gamma(B^0\rightarrow [K^-\pi^+]_DK^+\pi^-)/\Gamma(B^0\rightarrow [K^+\pi^-]_DK^+\pi^-)$ to be $(4.1 ^{+ 5.6 + 2.8}_{- 5.0 - 1.8}) \times 10^{-2}$,
and set an upper limit of ${\cal R}_{DK^{*0}} < 0.16$ at the 95\% confidence level.
\end{abstract}


\maketitle

\tighten

{\renewcommand{\thefootnote}{\fnsymbol{footnote}}}
\setcounter{footnote}{0}

Determination of the parameters of the standard model is important as a consistency check
and as a way to search for new physics.
In the standard model, the Cabibbo-Kobayashi-Maskawa (CKM) matrix~\cite{CKM} $V$ consists of four independent weak interaction parameters for the quark sector;
the three $CP$-violating phases $\phi_1$, $\phi_2$ and $\phi_3$ are defined as the angles of one particular CKM unitarity triangle
with the latter defined as $\phi_3 \equiv \arg{(-V_{ud}{V_{ub}}^*/V_{cd}{V_{cb}}^*)}$.
This phase is less accurately determined than the other two~\cite{phi12}.
In the usual quark-phase convention where large complex phases appear only in $V_{ub}$ and $V_{td}$~\cite{Wolfenstein},
the measurement of $\phi_3$ is equivalent to the extraction of the phase of $V_{ub}$ relative to the phases of other CKM matrix elements.
To date, the $\phi_3$ measurement has been advanced mainly by exploiting charged $B$ meson decays into $D^{(*)}K^+$ final states
~\cite{GLW_Belle_result, Dalitz_Belle_result, GLW_CDF_result, GLW_BaBar_result, ADS_BaBar_result, Dalitz_BaBar_result, ADS_CDF_result, ADS_Belle_result, LHCb_result}
wherein the $CP$ sensitivity is due to the interference between the two amplitudes of $\bar{D}^{(*)0}$ and $D^{(*)0}$ decays into a common final state.

\begin{figure}[ht]
  \begin{center}
    \leavevmode
    \subfigure
    {\includegraphics[width=0.21\textwidth]{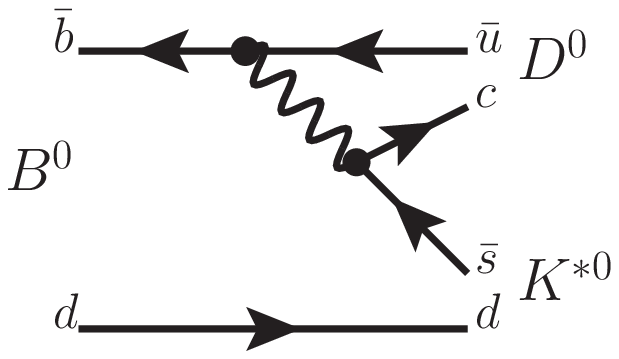}}
    \hspace{6mm}
    \subfigure
    {\includegraphics[width=0.21\textwidth]{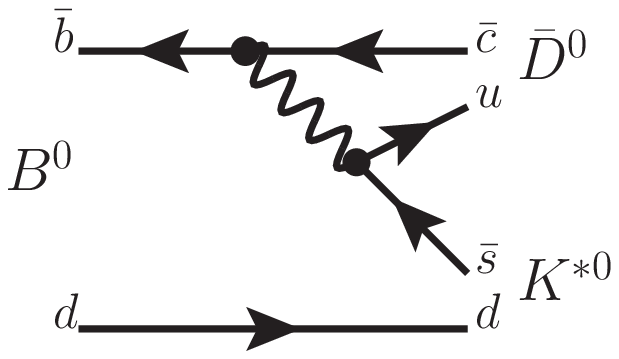}}
    \caption{Diagrams for the $B^0 \rightarrow D^0 K^{*0}$ and $B^0 \rightarrow \bar{D}^0 K^{*0}$ decays.
      The $\phi_3$ dependence in the $b\rightarrow u$ transition
      is extracted from the interference of the two decay paths,
      which occurs when the $\bar{D}^0$ and $D^0$ mesons decay to the same final state.}
    \label{fig:diag_dk}
  \end{center}
\end{figure}

In this paper, we consider the neutral meson decay $B^0 \rightarrow DK^{*0}$ as an alternative process for measuring the angle $\phi_3$.
As shown by the Feynman diagrams in Fig.~\ref{fig:diag_dk},
a weak decay of the $B$ meson is tagged by the $K^{*0}$ decaying into $K^+\pi^-$~\cite{charge}.
We measure the ratio ${\cal R}_{DK^{*0}}$~\cite{ADS, ADS_2} defined as
\begin{eqnarray}
  {\cal R}_{DK^{*0}} &\equiv& \frac{\Gamma(B^0\rightarrow [K^-\pi^+]_DK^+\pi^-)}
  {\Gamma(B^0\rightarrow [K^+\pi^-]_DK^+\pi^-)} \label{eq:rdk} \nonumber \\
  &=& r_S^2 + r_D^2 + 2kr_Sr_D \cos{(\delta_S+\delta_D)}\cos{\phi_3},
\end{eqnarray}
where $r_D \equiv |A(D^0\rightarrow K^+\pi^-)/A(D^0\rightarrow K^-\pi^+)|$ is the ratio for $D$ decay amplitudes
and $\delta_D$ is the strong phase difference of the two $D$ decays appearing in this ratio.
Both $r_D$ and $\delta_D$ have been obtained experimentally~\cite{HFAG}.
The parameters $r_S$, $\delta_S$ and $k$ are defined as
\begin{eqnarray}
  r_S^2 \equiv \frac{\Gamma(B^0\rightarrow D^0K^+\pi^-)}{\Gamma(B^0\rightarrow \bar{D}^0K^+\pi^-)} = \frac{\int dp A_{b\rightarrow u}^2(p)}{\int dp A_{b\rightarrow c}^2(p)}, \\
  k\mathrm{e}^{i\delta_S} \equiv \frac{\int dp A_{b\rightarrow c}(p)A_{b\rightarrow u}(p)\mathrm{e}^{i\delta(p)}}{\sqrt{\int dp A_{b\rightarrow c}^2(p) \int dp A_{b\rightarrow u}^2(p)}},
\end{eqnarray}
where $A_{b\rightarrow c}(p)$ and $A_{b\rightarrow u}(p)$ are the magnitudes of the amplitudes for the $b\rightarrow c$ and $b\rightarrow u$ transitions, respectively,
and $\delta(p)$ is the relative strong phase.
The variable $p$ indicates the position in the $DK^+\pi^-$ Dalitz plot.
In this analysis, we calculate the integrals over a phase space of the state $DK^*(892)^0$.
In the case of a two-body $B$ decay,
$r_S$ becomes the ratio of the amplitudes for $b\rightarrow u$ and $b\rightarrow c$ and $k$ becomes 1.
The value of $r_S$ is expected to be around 0.4, which is obtained from $|V_{ub}V^*_{cs}|/|V_{cb}V^*_{us}|$ and depends on strong interaction effects.
According to a simulation study using a Dalitz model based on recent measurements~\cite{BaBar_k},
the value of $k$ is around 0.95 in the phase space of interest here.
One observable ${\cal R}_{DK^{*0}}$ is not enough to extract the four unknowns $\phi_3$, $r_S$, $k$, and $\delta_S$.
However, the measurements for other $D$ decays such as $D\rightarrow K^+K^-$ and $K_S\pi^0$ provide additional information needed to extract $\phi_3$,
where the observable ${\cal R}_{DK^{*0}}$ should be defined in the same phase space of the $B^0$ decay between different $D$ decays so that the same parameters $r_S$, $k$, and $\delta_S$ can be used.
The decay in the numerator of Eq.~(\ref{eq:rdk}) is the signal mode, referred to as the ``suppressed mode,"
while the decay in the denominator is the calibration mode referred to as the ``favored mode."

This result is based on a data sample that contains 772~$\times 10^6~B\bar{B}$ pairs,
collected  with the Belle detector at the KEKB asymmetric-energy $e^+e^-$ (3.5 on 8~GeV) collider~\cite{KEKB} operating at the $\Upsilon(4S)$ resonance.
The Belle detector is a large-solid-angle magnetic spectrometer that consists of a silicon vertex detector,
a 50-layer central drift chamber (CDC),
an array of aerogel threshold Cherenkov counters (ACC),
a barrel-like arrangement of time-of-flight scintillation counters (TOF),
and an electromagnetic calorimeter comprised of CsI(Tl) crystals located inside a superconducting solenoid coil that provides a 1.5~T magnetic field.
An iron flux-return located outside of the coil is instrumented to detect $K_L^0$ mesons and to identify muons.
The detector is described in detail elsewhere~\cite{Belle}.

Charged kaon and pion candidates are identified using ionization loss in the CDC and information from the ACC and the TOF.
The efficiency is 85--95\% and the probability of misidentification is 10--20\%.
We reconstruct $D$ mesons from pairs of oppositely-charged kaon and pion candidates.
We require that the invariant mass is within $\pm$15~${\rm MeV}/c^2$ ($\pm3\sigma$) of the nominal $D^0$ mass.
$K^{*0}$ candidates are reconstructed from $K^+\pi^-$ pairs.
We require that the invariant mass is within $\pm$50~${\rm MeV}/c^2$ of the nominal $K^{*0}$ mass.
We combine $D$ and $K^{*0}$ candidates to form $B^0$ mesons.
Candidate events are identified by the energy difference $\Delta E \equiv \sum_{i}E_i - E_{\mathrm b}$
and the beam-constrained mass $M_{\rm bc} \equiv \sqrt{E_{\mathrm b}^2 - |\sum_{i}\vec{p}_i|^2}$,
where $E_{\mathrm b}$ is the beam energy and $\vec{p}_i$ and $E_i$ are the momenta and energies, respectively,
of the $B^0$ meson decay products in the $e^+e^-$ center-of-mass (CM) frame.
We select events with $5.271~{\rm GeV}/c^2 < M_{\rm bc} < 5.287~{\rm GeV}/c^2$ and $-0.1~{\rm GeV} < \Delta E < 0.3~{\rm GeV}$.
In the rare case where there are multiple candidates in an event,
the candidate with $M_{\rm bc}$ closest to its nominal value is chosen.

Among other $B$ decays, the most serious background for the suppressed mode comes from $\bar{B}^0\rightarrow [\bar{K}^{*0}K^+]_{D^+}\pi^-$.
This decay produces the same final state as the $B^0\rightarrow DK^{*0}$ signal,
and the product branching fraction is about 10 times higher than that expected for the signal.
To suppress this background, we exclude candidates for which the invariant mass of the $K^- \pi^+ K^+$ system is within $\pm$18~${\rm MeV}/c^2$ ($\pm3\sigma$) of the nominal $D^+$ mass.
The relative loss in the signal efficiency is 0.5\%.

Large combinatorial background of true $D^0$ and random $K^+$ and $\pi^-$ combinations from the $e^+e^- \rightarrow c\bar{c}$ process and other $B\bar{B}$ decays
is reduced if the $D^0$ is a decay product of $D^{*+} \rightarrow D^0 \pi^+$
by using the mass difference $\Delta M$ between the $[K^- \pi^+]_D \pi^+$ and $[K^- \pi^+]_D$ systems,
where a $\pi^+$ candidate is added to the latter to form the former.
If $\Delta M > 0.15~{\rm GeV}/c^2$ for any additional $\pi^+$ candidate not used in the $B$ candidate reconstruction, the event is retained.
This requirement removes 24\% of $c\bar{c}$ background and 14\% of $B\bar{B}$ background according to Monte Carlo (MC) simulation.
The relative loss in signal efficiency is 5.0\%.

To discriminate the large combinatorial background dominated by the two-jet-like $e^+e^-\rightarrow q\bar{q}$ continuum process,
where $q$ indicates $u$, $d$, $s$ or $c$, a multivariate analysis is performed using the following nine variables.
1) A variable obtained from the Fisher discriminants based on modified Fox-Wolfram moments~\cite{SFW}
where the coefficients of the Fisher discriminants are optimized using the signal and $q\bar{q}$ MC samples.
This variable exploits the event topology,
which is spherical and jet-like for $B\bar{B}$ and $q\bar{q}$ events, respectively.
2) The angle in the CM frame between the thrust axes of the $B$ decay and the detected remainders.
For the latter, we assign the pion mass to all the charged particles
and use photons with energy above 0.1~${\rm GeV}$.
3) The signed difference of the vertices between the $B$ candidate and the remaining charged tracks.
For the signal event, the absolute value tends to be larger because of the longer lifetime of the $B$ meson.
4) The angle between the $K$ candidate from the $D$ decay and the $B$ candidate in the rest frame of the $D$ candidate.
Its distribution is flat for signal events but peaked near the extreme values for $q\bar{q}$ background.
5) The expected flavor dilution factor described in Ref.~\cite{TaggingNIM}.
It ranges from zero for no flavor information to unity for unambiguous flavor assignment.
$B$ candidates tend to have a larger flavor dilution factor than $q\bar{q}$ background.
6) The angle $\theta$ between the $B$ meson momentum direction and the beam axis in the CM frame.
The $B$ decays follow a $1 - \cos^2 \theta$ distribution, while the $q\bar{q}$ background is nearly flat in $\cos \theta$.
7) The distance of closest approach between the trajectories of the $K^{*}$ and $D$ candidates.
The value is close to zero for the signal but tends to be larger for the $c\bar{c}$ background.
8) The difference between the sum of the particle charges in the $D$ hemisphere and the sum in the opposite hemisphere,
excluding those used in the reconstruction of the $B$ meson.
The average charge difference is 0 for the signal events but $\pm 4/3$ for the $c\bar{c}$ events, depending on the flavor of the $B$ candidate.
9) The angle between the $D$ and $\Upsilon(4S)$ directions in the rest frame of the $B$ candidate.
The cosine distribution is about flat for signal events but peaks toward $+1$ for $c\bar{c}$ events.

To effectively combine these nine variables,
we employ the NeuroBayes neural network package \cite{NB}.
The NeuroBayes output is denoted as $C_{\rm NB}$ with a range of [$-$1, 1].
For example, events at $C_{\rm NB} \sim 1$ are signal-like and events at $C_{\rm NB} \sim -1$ are $q\bar{q}$-like.
The training for the neural network optimization is performed by using the signal and the $q\bar{q}$ MC samples,
each of which contains 100,000~events after the event-selection requirements.
For the latter sample, we loosen the requirement on $M_{\rm bc}$ to $5.23~{\rm GeV}/c^2 < M_{\rm bc} < 5.27~{\rm GeV}/c^2$ to obtain a larger number of events,
since all the input parameters have little correlation with $M_{\rm bc}$.

The $C_{\rm NB}$ distribution peaks at $|C_{\rm NB}|\sim 1$ and is therefore difficult to represent with a simple analytic function.
However, the transformed variable
\begin{eqnarray}
  C'_{\rm NB} &=& \ln \frac{C_{\rm NB}-{C_{{\rm NB,} {\rm low}}}}{C_{{\rm NB,} {\rm high}}-C_{\rm NB}}\ ,
\end{eqnarray}
where $C_{{\rm NB,} {\rm low}} = -0.6$ and $C_{{\rm NB,} {\rm high}} = 1.0$,
has a distribution that can be modelled by a Gaussian.
The events with $C_{\rm NB} < -0.6$ are rejected.
The background rejection rate is 70.5\%, while the signal loss is 3.9\%.

The number of signal events is obtained by a two-dimensional unbinned extended maximum likelihood fit to $\Delta E$ and $C'_{\rm NB}$.
The fits are applied separately for favored and suppressed modes.
For both modes, we categorize five common contributions.
These are the $DK^{*0}$ signal,
the $\bar{D}^0\rho^0$ background,
the combinatorial $B\bar{B}$ background,
the $q\bar{q}$ background,
and 
the backgrounds that have peaks in the signal region of $\Delta E$ and $C'_{\rm NB}$ (``peaking background'').
In the favored mode, we include two more components: $\bar{D}^0K^+$ and $\bar{D}^0\pi^+$.
The $B^0\rightarrow \bar{D}^0\rho^0$ decay satisfies the selection criteria when a pion from the $\rho^0$ decay is misidentified as a kaon.
This component also includes other decays that satisfy the selection criteria when a pion in the final state is misidentified as a kaon.
The peaking background for the suppressed mode consists of $B^0\rightarrow [K^-\pi^+\pi^-]_{D^-}K^+$ and $B^0\rightarrow [K^+K^-]_{D^0}\pi^+\pi^-$
while the peaking background for the favored mode consists of $B^0\rightarrow [K^+\pi^-\pi^-]_{D^-}K^+$.
For the $B^0\rightarrow \bar{D}^0K^+$ and $\bar{D}^0\pi^+$ backgrounds,
a pion candidate is added to reconstruct $K^{*0}$,
where the latter satisfies the selection when the $\pi^+$ is misidentified as $K^+$.
We prepare two-dimensional probability density functions (PDFs) for each component as a product of one-dimensional PDFs on $\Delta E$ and $C'_{\rm NB}$,
since the correlation between $\Delta E$ and $C'_{\rm NB}$ is found to be small.

The $\Delta E$ PDFs for a favored mode are parameterized by a double Gaussian for signal, a double Gaussian for $\bar{D}^0\rho^0$,
an exponential function for $B\bar{B}$ background,
a linear function for $q\bar{q}$ background,
a Crystal Ball function for $\bar{D}^0K^+$,
and a double bifurcated Gaussian for $\bar{D}^0\pi^+$.
The means and widths of the double-Gaussian PDFs for the signal and $\bar{D}^0\rho^0$ components are fixed from MC samples.
The mean of the $\Delta E$ distribution for $\bar{D}^0\rho^0$ is higher than that for the signal by about 70~${\rm MeV}$ due to misidentification of a pion as a kaon.
The parameters of the exponential and linear PDFs are allowed to float.
The $\Delta E$ PDF for the peaking background is defined to be that of the signal,
and the yield is fixed by the world-average value of the branching fraction~\cite{PDG}.
The mean values of $\Delta E$ for $\bar{D}^0K^+$ and $\bar{D}^0\pi^+$ are higher than those for the signal due to one additional pion and a misidentification for the latter mode.
The shape parameters of the $\Delta E$ PDFs for these components are determined from MC
and their yields are fixed by the world-average value of the branching fraction.

The $C'_{\rm NB}$ PDF is a sum of two Gaussians for each component.
The shapes for the signal and $B\bar{B}$ background are fixed from the MC samples of each decay model.
The $C'_{\rm NB}$ PDF for $\bar{D}^0\rho^0$ is defined by the same function as that of $B\bar{B}$ background.
The $C'_{\rm NB}$ PDF for the peaking background is described 
as a weighted sum of MC-based PDFs for all the constituents.
The shape for the $q\bar{q}$ background is fixed from the $M_{\rm bc}$ sideband data sample in data defined by $5.23~{\rm GeV}/c^2 < M_{\rm bc} < 5.27~{\rm GeV}/c^2$.
The validity of this use of the $M_{\rm bc}$ sideband sample,
which is reasonable since all inputs for $C'_{\rm NB}$ have little correlation with $M_{\rm bc}$,
is checked using MC samples.
The $C'_{\rm NB}$ PDF for $\bar{D}^0K^+$ and $\bar{D}\pi^+$ is the same as that of the $B\bar{B}$ background.

\begin{figure}[htbp]
  \begin{center}
    \leavevmode
    {\includegraphics[width=0.5\textwidth]{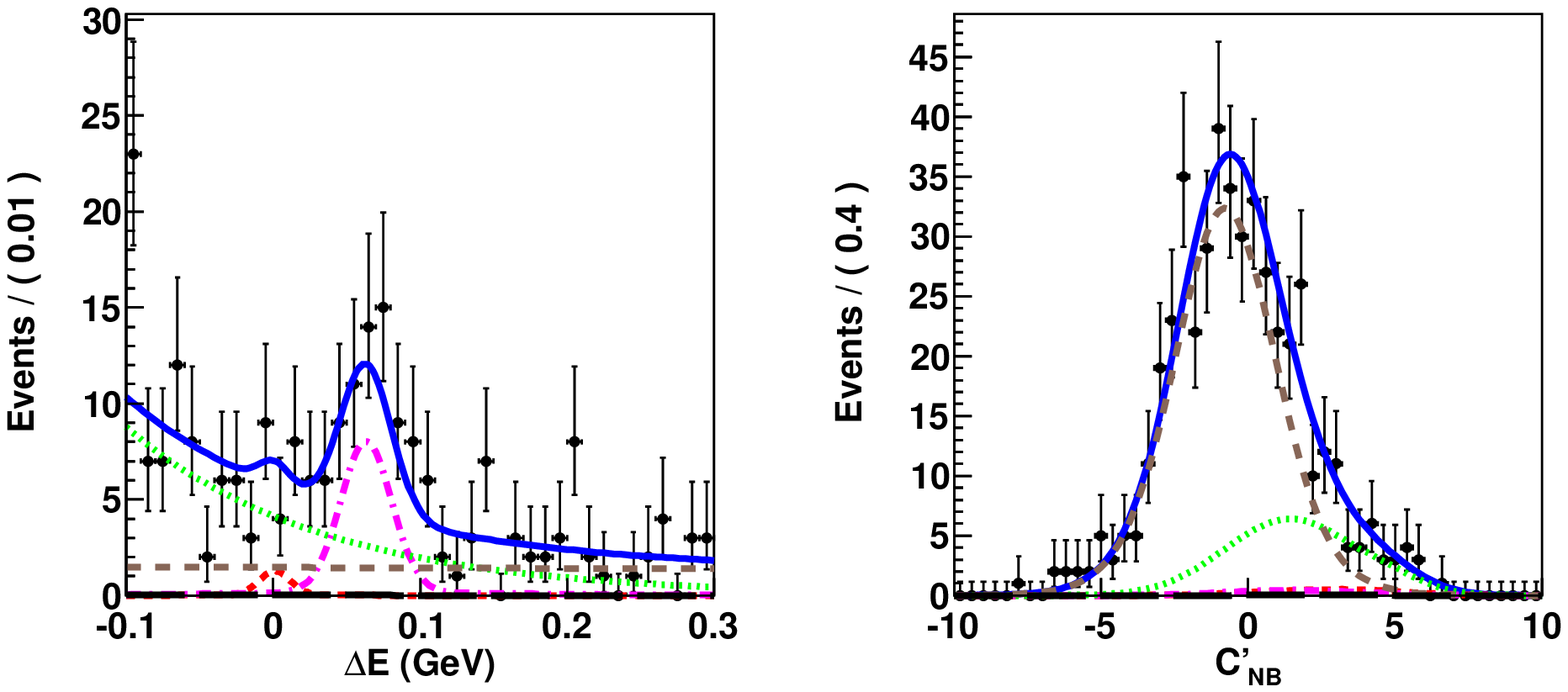}}
    \hspace{3mm}
    {\includegraphics[width=0.5\textwidth]{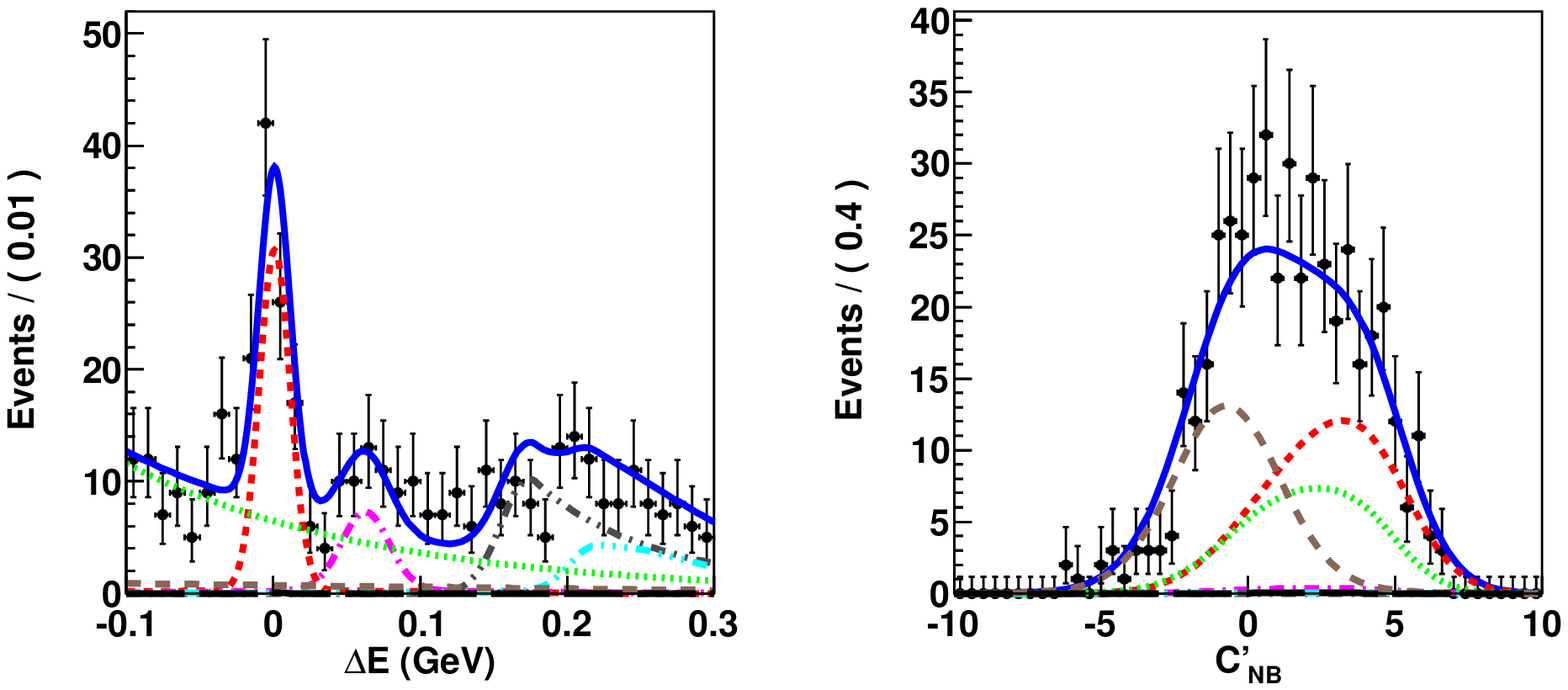}}
    \caption{The projections of the fits to data for the suppressed mode~(upper) and the favored mode~(lower):
      the $\Delta E$ projection for $3 < C'_{\rm NB} < 10$~(left) and the $C'_{\rm NB}$ projection for $|\Delta E|<0.03\,{\rm GeV}$~(right).
      The fitted data samples are shown by the dots with error bars and the total PDFs are shown by the solid blue curve.
      Individual components are shown by
      the dashed red ($DK^{*0}$ signal),
      the dash-dotted magenta ($\bar{D}^0\rho^0$),
      the short dashed green (combinatorial $B\bar{B}$ background),
      the long dashed brown ($q\bar{q}$ background),
      the very long dashed black (peaking backgrounds),
      the dash-dot-dotted gray ($\bar{D}^0K^+$),
      and the dash-dot-dot-dotted aqua ($\bar{D}^0\pi^+$).
    }
    \label{fig:fit}
  \end{center}
\end{figure}

The results of the fits for suppressed and favored modes are shown in Fig.~\ref{fig:fit} and presented in Table~\ref{table:summary}.
We obtain the ratio ${\cal R}_{DK^{*0}}$ to be 
\begin{eqnarray}
  {\cal R}_{DK^{*0}} &=& \frac{N_{\rm sup}/\epsilon_{\rm sup}}{N_{\rm fav}/\epsilon_{\rm fav}} \nonumber \\
  &=& (4.1 ^{+ 5.6 + 2.8}_{- 5.0 - 1.8}) \times 10^{-2}, \nonumber
\end{eqnarray}
where $N_{\rm sup\ (fav)}$ is the signal yield for the suppressed (favored) mode, $\epsilon_{\rm sup\ (fav)}$ is the detection efficiency obtained from a MC study for the suppressed (favored) mode.

We list the sources of systematic uncertainties in Table~\ref{table:syst_r}.
The uncertainties of the PDF shape parameters are estimated by varying the determined parameters of the PDFs by $\pm1\sigma$.
The uncertainties due to the $C'_{\rm NB}$ PDFs for $\bar{D}^0\rho^0$, combinatorial $B\bar{B}$, $\bar{D}^0K^+$, and $\bar{D}^0\pi^+$ are estimated by replacing their PDFs with the signal PDF.
The uncertainty due to the PDF shape for $q\bar{q}$ is the largest systematic uncertainty.
The uncertainty due to the yields of the peaking background is conservatively estimated by applying 0 and 2 times the nominal expected yields.
The systematic uncertainty associated with the peaking background is small because of its small expected yield.
The uncertainty due to the yields of $\bar{D}^0K^+$ and $\bar{D}^0\pi^+$ is estimated by taking into account the uncertainty of the efficiencies and the branching fractions.
We check the fit bias by generating 10,000 pseudo-experiments for each of the suppressed and favored modes.
We obtain an almost standard Gaussian distribution for the pull,
and take the product of the mean of the pull and the error of the nominal fit.
MC statistics and the uncertainties in the efficiencies of particle identification dominate the systematic uncertainty in detection efficiency.
The uncertainties in the efficiencies of particle identifications are determined from the decay $D^{*+}\rightarrow D^0\pi^+$
followed by $D^0\rightarrow K^-\pi^+$.
The uncertainty due to the charmless $B^0\rightarrow K^{*0}K^+\pi^-$ decay is obtained from the upper limit of its branching ratio~\cite{PDG}
and the efficiency estimated by assuming a non-resonant distribution in phase space.
The uncertainties due to the favored mode are estimated in a similar manner as for the suppressed mode and are found to be small.

The distribution of the likelihood $\mathcal{L}$ is obtained by convolving the likelihood in the ($\Delta E, C'_{\rm NB}$) two-dimensional fit and an asymmetric Gaussian whose widths are the negative and positive systematic errors.
We set a 95\% confidence level (C.L.) upper limit for ${\cal R}_{DK^{*0}}$ to be ${\cal R}_{DK^{*0}} < 0.16$.
We obtain an upper limit of $r_S < 0.4$, corresponding to 95\% C.L. limit of $R_{DK^{*0}}$,
by conservatively assuming that $r_S$ is much larger than $r_D$ so that $R_{DK^{*0}} = r_S^2$.
The uncertainties due to the signal yield of the favored mode are found to be negligible.

\begin{table}[tbp]
  \caption{Summary of the results. 
    The errors for $N$ and ${\cal R}_{DK^{*0}}$ are statistical only.}
  \label{table:summary}
  \begin{center}
    \begin{tabular}{cccc}
      \hline \hline
      Mode & $\epsilon$ (\%) & $N$ & ${\cal R}_{DK^{*0}}$ \\ \hline
      $B^0 \rightarrow [K^+\pi^-]_D K^{*0}$ & $21.0\pm0.3$ & $190^{+ 22.3}_{- 21.2}$ & 
      \multirow{2}{*}{($4.1^{+ 5.6}_{- 5.0}$)$\times10^{-2}$} \\
      $B^0 \rightarrow [K^-\pi^+]_D K^{*0}$ & $20.9\pm0.3$ & $7.7^{+ 10.6}_{- 9.5}$ & \\
      \hline \hline
    \end{tabular}
  \end{center}
\end{table}

\begin{table}[tbp]
  \caption{Summary of the systematic uncertainties for ${\cal R}_{DK^{*0}}$.
  }
  \label{table:syst_r}
  \begin{center}
    \begin{tabular}{lccc}
      \hline \hline
      {\footnotesize Source~ }&{\footnotesize  ~~~Uncertainty [$10^{-2}$]~~~ } \\
      \hline
      {\footnotesize Signal PDFs                          }&{\footnotesize  $+ 0.1 - 0.2$ } \\
      {\footnotesize $\bar{D}^0\rho^0$ PDFs               }&{\footnotesize  $+ 0.0 - 0.1$ } \\
      {\footnotesize Combinatorial $B\bar{B}$ PDFs        }&{\footnotesize  $+ 1.8 - 1.2$ } \\
      {\footnotesize Peaking background PDFs              }&{\footnotesize  $+ 0.1 - 0.1$ } \\
      {\footnotesize $q\bar{q}$ PDFs                      }&{\footnotesize  $+ 2.2 - 1.4$ } \\
      {\footnotesize $\bar{D}^0K^+$ PDFs                  }&{\footnotesize  $+ 0.0 - 0.0$ } \\
      {\footnotesize $\bar{D}^0\pi^+$ PDFs                }&{\footnotesize  $+ 0.0 - 0.1$ } \\
      {\footnotesize Fit bias                             }&{\footnotesize  $+ 0.4 - 0.0$ } \\
      {\footnotesize Efficiency                           }&{\footnotesize  $+ 0.1 - 0.1$ } \\
      {\footnotesize Charmless decay                      }&{\footnotesize  $+ 0.0 - 0.3$ } \\
      \hline
      {\footnotesize Total                                }&{\footnotesize  $+ 2.8 - 1.8$ } \\
      \hline \hline
    \end{tabular}
  \end{center}
\end{table}

In summary, we report a result of the measurement of the ratio ${\cal R}_{DK^{*0}}$, using a 711~${\rm fb}^{-1}$ data sample collected by the Belle detector.
We obtain ${\cal R}_{DK^{*0}} = (4.1^{+ 5.6 + 2.8}_{- 5.0 - 1.8})\times 10^{-2}$, which can be used to extract $\phi_3$ by combining with other observables related to the same dynamical parameters $r_S$, $\delta_S$ and $k$.
Since the value of ${\cal R}_{DK^{*0}}$ is not significant,
we set an upper limit of ${\cal R}_{DK^{*0}} < 0.16$ (95\% C.L.); this is the most stringent limit to date.
Possible reasons for the small $r_S$ are destructive interference between the two $D$ decays,
destructive interference between $DK^{*0}$ and other $DK^+\pi^-$ states,
or a small ratio of magnitudes of amplitudes for $B^0\rightarrow D^0K^{*0}$ over $B^0\rightarrow \bar{D}^0K^{*0}$.

We thank the KEKB group for excellent operation of the
accelerator; the KEK cryogenics group for efficient solenoid
operations; and the KEK computer group, the NII, and 
PNNL/EMSL for valuable computing and SINET4 network support.  
We acknowledge support from MEXT, JSPS and Nagoya's TLPRC (Japan);
ARC and DIISR (Australia); NSFC (China); MSMT (Czechia);
DST (India); INFN (Italy); MEST, NRF, GSDC of KISTI, and WCU (Korea); 
MNiSW (Poland); MES and RFAAE (Russia); ARRS (Slovenia); 
SNSF (Switzerland); NSC and MOE (Taiwan); and DOE and NSF (USA).


\begin{thebibliography}{99}
\bibitem{phi12}
  The other unitarity-triangle angles
  are defined as $\phi_1 \equiv \arg{(-V_{cd}{V_{cb}}^*/V_{td}{V_{tb}}^*)}$ and $\phi_2 \equiv \arg{(-V_{td}{V_{tb}}^*/V_{ud}{V_{ub}}^*)}$.
\bibitem{charge}
  The charge-conjugate modes are implicitly included unless otherwise stated.
\end{thebibliography}
\end{document}